\documentclass[aip,twocolumn,groupedaddress]{revtex4}
\usepackage{graphicx}
\usepackage{bm}
\usepackage{times}
\usepackage{amsmath}
\usepackage{float}

\begin{document} 

\title{High-dynamic-range transmission-mode detection of synchrotron radiation using X-ray excited optical luminescence in diamond\footnote{accepted for publication in J. Synchrotron Rad.}}

\author{Stanislav Stoupin}
\email{sstoupin@cornell.edu} 
\affiliation{Cornell High Energy Synchrotron Source, Cornell University, Ithaca, NY 14853, USA}

\author{Sergey Antipov}
\affiliation{Euclid Techlabs LLC, Solon, OH 44139, USA} 

\author{Alexander M. Zaitsev}
\affiliation{College of Staten Island and Graduate School of the City University of New York, Staten Island, NY 10314, USA and Gemological Institute of America, New York, NY 10036, USA}

\begin{abstract}
We demonstrate enhancement of X-ray excited optical luminescence in a 100-micron-thick diamond plate by introduction of defect states via electron beam irradiation and subsequent high-temperature annealing. The resulting X-ray transmission-mode scintillator features a linear response to incident photon flux in the range of 7.6$\times$10$^8$ to 1.26$\times10^{12}$ photons/s/mm$^2$ for hard X-rays (15.9~keV) using exposure times from 0.01 to 5~s. These characteristics enable a real-time transmission-mode imaging of X-ray photon flux density without disruption of X-ray instrument operation.
\end{abstract}

\maketitle 

\section{Introduction}
Noninvasive (or minimally invasive) visualization of X-ray beam profiles under in-operando conditions is of high practical importance at experimental stations of large-scale synchrotron and XFEL user facilities. The desired generalized device functionality can be described as real-time imaging of flux density distribution in a chosen observation plane placed across the direction of propagation of X-ray beams. Common metrics, which reflect performance of the experimental station, such as, average beam position, intensity/photon flux over a chosen region of interest as well as position and intensity fluctuations can be derived from the real-time profiles of flux density. 
A recently demonstrated, quantitative approach for flux density monitoring features detection of electrical charge in a lithographically patterned (pixelated) diamond plate \cite{Zhou15}. X-ray transmission-mode scintillators (producing X-ray excited optical luminescence) with low X-ray absorption, coupled to a visible-light area detectors is an alternative, more straightforward strategy towards implementation of the imaging functionality. Diamond is a preferable choice for the X-ray transmission-mode scintillator due to its low X-ray absorption, high radiation hardness, remarkable thermal and mechanical properties. Video-monitoring of X-ray excited optical luminescence in thin diamond plates is commonly used for this purpose, however, predominantly in a semi-quantitative manner, where the observed profile is evaluated based on visual appearance characteristics (e.g., spatial resolution, relative brightness) often without documented knowledge on response linearity and dynamic detection range. Such advanced characterization is perhaps unnecessary for monitoring profiles of intense synchrotron beams upstream of the beamline monochromator (beamline front-end) \cite{Degenhardt13,Kosciuk16,Takahashi16}. In other cases (e.g., X-ray beams downstream of the monochromator), a more quantitative approach is required. 
Park et al. \cite{Park18} recently demonstrated $\simeq$~3~$\mu$m beam position stability for a monochromatic beam using a commercial diamond screen as a real-time imaging detector. However, no information on the response linearity is provided, therefore the dynamic detection range of their study remains unclear. 
In this work, we achieved enhancement of X-ray optical luminescence in a thin diamond plate by introduction of additional defects via electron irradiation and subsequent high-temperature annealing. The resulting responsivity and the dynamic range of the luminescence detection was increased by more than one order of magnitude compared to commercially available diamond screens of the same thickness. Contrary to the commercial diamond screens, the response to the incident X-ray photon flux was found to be linear in the range of 7.6$\times$10$^8$ to 1.26$\times10^{12}$ photons/s/mm$^2$.

\section{Samples}
The samples procured for this study were circular polycrystalline diamond plates prepared using chemical vapor deposition (CVD) with diameter of 10~mm and thickness of 100 micron. These were of the nominal “tool”, “thermal” and “optical” grades (supplier specification), presumably of different impurity concentration. An additional polycrystalline diamond plate of the "optical" grade of the same shape was acquired from the same supplier. It was subjected to irradiation and subsequent annealing to generate luminescent centers (irradiated and annealed plate). To create vacancies in the diamond lattice irradiation was performed with an energetic electron beam. Practically any diamond contains impurities. Both diamond plates of the "optical" grade had completely transparent visual appearance. The main impurity was N with expected concentration on the order of 0.1-10 ppm. Annealing was performed in vacuum to promote formation of NV (nitrogen-vacancy) defect centers.
In this work we made no attempts to study boron doped diamond samples for the following reasons. While it is known that some natural and synthetic boron-doped single crystals produce enhanced luminescence the characteristic luminescence lifetimes are on the order of seconds \cite{Eaton-Magana11,Gaillou12}, which can result in a nonlinear response for real-time detection. Prior studies indicate that boron is not an efficient luminescent center in boron-doped polycrystalline CVD diamond \cite{Graham94,Iakoubovskii00-prb}. 

\begin{figure}
\centering\includegraphics[width=0.5\textwidth]{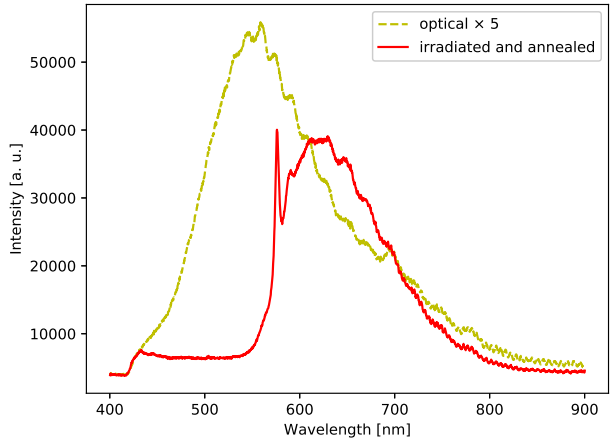}
\caption{Spectra of UV-excited luminescence in the irradiated and annealed diamond plate (red solid line) and 
diamond plate of optical grade (yellow dashed line; data collected using 5 times longer exposure time per point).}
\label{fig:UV}
\end{figure}

Luminescence spectra under UV excitation at wavelength of 360 nm were measured for all samples. Selected spectra are shown in Fig.~\ref{fig:UV}. The luminescence spectrum of the unannealed "optical" grade sample (from here on referred to as "optical") is shown with yellow dashed line. It can be described as a broad band with a maximum response in the spectral range from 500 to 600~nm. This band is due to characteristic optically active defects, which usually dominate luminescence of as-grown nitrogen-doped CVD diamonds. In luminescence measured at low temperature, these optical centers produce zero-phonon line at wavelength 468 nm (not clearly seen in the present spectrum taken at room temperature) \cite{Zaitsev_book, Iakoubovskii00}.
Although the 486 nm center is very typical for nitrogen-doped CVD diamonds, its origin has not been identified yet. The luminesence spectrum of the "thermal" grade sample was found to be relatively weak, having similar spectral shape (not shown in the figure). The sample of "tool" grade did not produce any reasonably measurable luminescence. The spectral response of the irradiated and annealed sample (shown with red solid line) was much stronger compared to that of the "optical" grade sample. Its appearance is characteristic to the NV0 center \cite{Zaitsev_book}, which includes the zero-phonon line (narrow peak at 575 nm) and the broad phonon side band above it.

\section{Experiment, results and discussion}
The characterization experiment was conducted at 2B beamline of Cornell Synchrotron Radiation Source (Cornell University, USA). The beamline features CHESS compact undulator source and a side-bounce monochromator operating at a set of fixed photon energies (9.7, 15.9, 18.65, 22.5~keV). 
Diamond screens were placed in light-tight environment at 45~degrees with respect to incident X-ray beam, which was shaped to 1$\times$1~mm$^2$ size using X-ray slits. 
X-ray excited luminescence was measured using Mako G319C camera (Allied Vision) equipped with an objective lens. The camera was placed at a distance of about 100~mm from the sample. An ionization chamber (IC0) was placed upstream of the diamond screen to monitor the incident photon flux and another ionization chamber (IC1) was set downstream the diamond to monitor photon flux transmitted through the sample. The experimental setup is shown schematically in Fig. 1 (top view). 

\begin{figure}
\centering\includegraphics[width=0.45\textwidth]{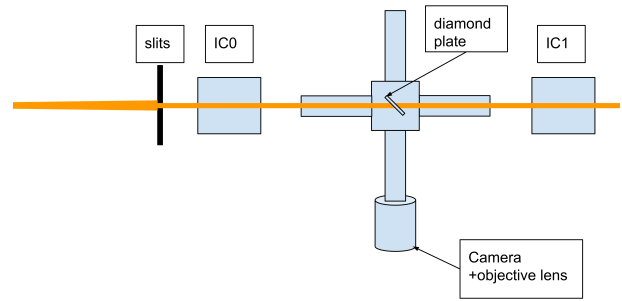}
\caption{Experimental setup (top view, see text for details).}
\label{fig:setup}
\end{figure} 

\begin{figure}
\centering\includegraphics[width=0.5\textwidth]{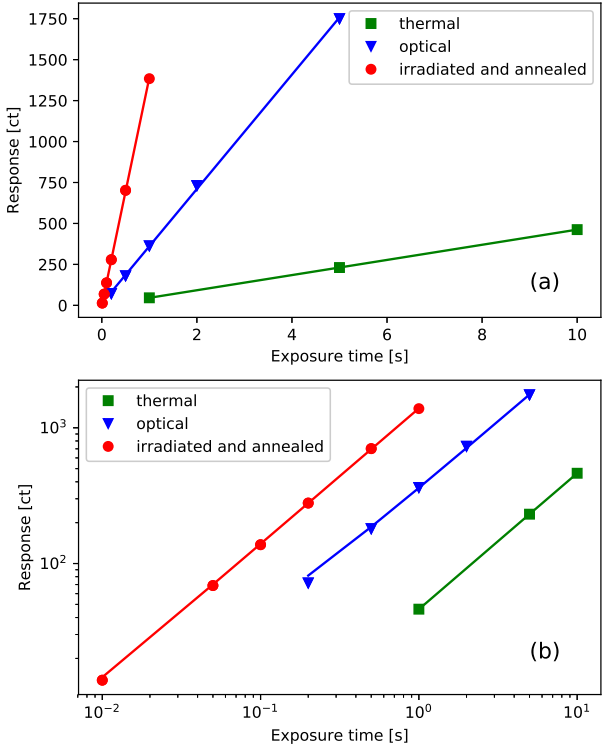}
\caption{X-ray excited optical luminescence (average response over the 1$\times$1~mm$^2$ X-ray beam footprint) as a function of exposure time for the as-received diamond plates (thermal and optical grades) and for the irradiated and annealed plate. The linear-linear plot (a) is supplemented with log-log plot (b). The solid lines are fits to the linear function.}
\label{fig:r_t}
\end{figure}

During preliminary tests with X-ray beam, luminescence intensity of the “tool” grade sample was found negligible compared to that of other samples. Therefore, results for this sample were excluded from analysis. This can be easily explained by dark appearance of the sample (high self-absorption of luminescence). 
The X-ray transmissivity was measured at 15.9 keV. The observed values were 0.965, 0.969 and 0.973 for the thermal, optical and the irradiated plates, respectively.  For a 100-micron-thick diamond plate oriented at 45 degrees with respect to the incident beam the transmission factors calculated using tabulated values for the mass attenuation coefficient and the mass energy-absorption coefficient \cite{nist_126} are 0.965 and 0.977, respectively. Thus, X-ray attenuation in the diamond plates was small as expected. 
In the first experiment, linearity of the system was explored by performing measurements at a fixed incident photon flux at 15.9 keV using different exposure times. The range of exposure times was from 0.01 up to 5~s (except a few 10 s exposures for the weakly luminescent "thermal" grade sample). This range could be described as real-time conditions for most experiments performed at synchrotron sources (excluding fast time-resolved experiments). The response was measured as a sum of pixel intensity across the image of the beam footprint, normalized by the number of pixels. An offset representing camera dark current (sum of pixel intensity for a region of the same size, not exposed to X-rays) was subtracted. The fluctuation of the response was found to be less than one percent. In the figures that follow the related uncertainties are less than the size of the figure markers. The values of the photon flux were evaluated using ion chamber flux calculator for N2 filled chambers of length 6 cm \cite{chess_ic_calc}. 
The response as a function of exposure time was found to be linear to within experimental uncertainties for all studied diamond plates. The results are shown in Fig.~\ref{fig:r_t}.

\begin{figure}
\centering\includegraphics[width=0.5\textwidth]{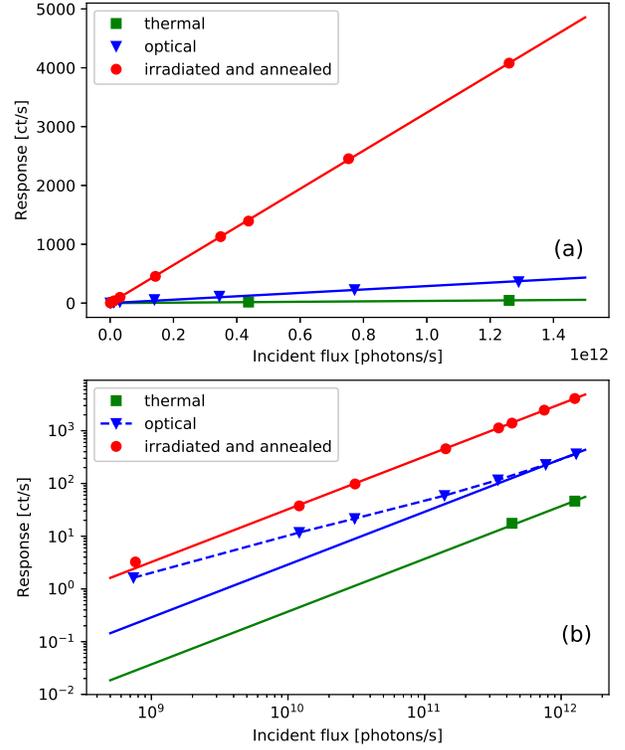}
\caption{X-ray excited optical luminescence (average time-normalized response over the 1$\times$1~mm$^2$ X-ray beam footprint) as a function of the incident photon flux for the as-received diamond plates (thermal and optical grades) and for the irradiated and annealed plate. 
The linear-linear plot (a) is supplemented with log-log plot (b). The dashed line connecting data for the optical grade sample in (b) illustrates nonlinear behavior. The solid lines are fits to the linear function.}
\label{fig:r_ft}
\end{figure}

This observation enabled further characterization of the luminescence as a function of the incident photon flux regardless of exposure time (normalization of the response by the exposure time was performed). Since the dynamic range of the camera was only 12 bit (up to 4096 counts per pixel), variable exposure time enabled greater total dynamic range for quantitative evaluation of luminescence intensity. 
In the second experiment the luminescence was evaluated at variable incident flux levels at a photon energy of 15.9 keV.
The time-normalized response (response from here on) as a function of the incident photon flux for the different diamond plates is shown in Fig.~~\ref{fig:r_ft}. Fits with a linear function are shown by the solid lines. The proportionality coefficients in the linear fits were 3.7$\times 10^{-11}$, 2.9$\times 10^{-10}$, and 3.2$\times 10^{-9}$ for the thermal, optical, and irradiated and annealed samples, respectively. For the sample of thermal grade, only two data points were measured, which precludes analysis of response linearity. The response of the optical grade sample deviates from the linear behavior (as shown by the dashed line), which reduces the dynamic range for detection of photon flux, while the response of the irradiated and annealed sample is linear to a good approximation. The response of the irradiated sample is greater by more than one order of magnitude. The resulting measured dynamic detection range is more than 3 orders of magnitude. The detected levels of the photon flux are from 7.6$\times 10^8$ to 1.26$\times 10^{12}$ photons/s. 

\begin{figure}
\centering\includegraphics[width=0.45\textwidth]{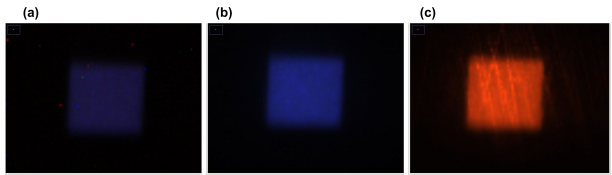}
\caption{Color images of X-ray excited optical luminescence in diamond plates of different grades: thermal (a), optical (b), irradiated and annealed sample (c).}
\label{fig:r_clr}
\end{figure}

Figure~\ref{fig:r_clr} shows color images (RGB) of the X-ray beam footprint taken at 15.9 keV. Different exposure times and flux levels were used to optimize image statistics. 

%\begin{figure}[h]
%\centering\includegraphics[width=0.75\textwidth]{fresnel.eps}
%\caption{Grazing incidence geometry illustrating the incident, the transmitted and the reflected wave. The transmitted wave generates ionization events in the mirror material due to X-ray absorption. The escaping fast photoelectons (light blue) ionize the surrounding gas while the low-energy secondary photoelectrons (gray) do not participate in this ionization process.}
%\label{fig:gzi}
%\end{figure} 

For the samples of thermal and optical grades the luminescence color is predominantly in the blue range. We note that the luminescence spectrum measured under UV excitation (Fig.~\ref{fig:UV}) corresponds to green color. This discrepancy suggests non-equivalence of X-ray and UV excitation. A more detailed interpretation of this observation falls outside the scope of the present study. For the irradiated and annealed sample the luminescence color is predominantly red due to the characteristic NV0 luminescence (consistent with the UV-excited luminescence spectrum). Some structure in luminescence of the irradiated sample was observed (lines of intensity propagating outside of the X-ray illuminated region). We attribute this effect to surface quality (luminescence re-scattering on surface defects/scratches). 

\begin{figure}
\centering\includegraphics[width=0.5\textwidth]{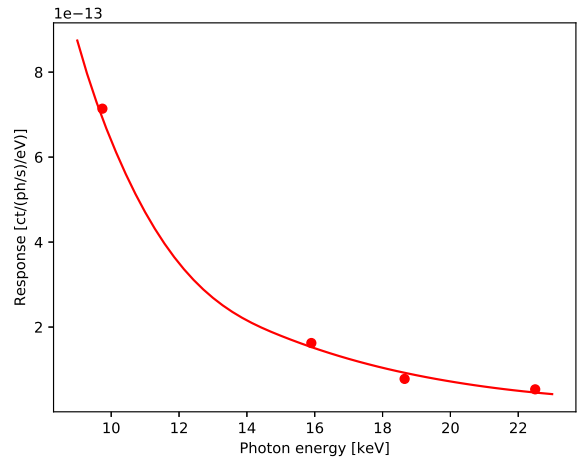}
\caption{X-ray excited optical luminescence (average energy-normalized response over the 1$\times$1~mm$^2$ X-ray beam footprint) as a function of the X-ray photon energy for the irradiated and annealed diamond plate. The solid line represents scaled absorptivity of X-rays in the 100-$\mu$m-thick diamond plate (Eq.~\ref{eq:R}).}
\label{fig:r_en}
\end{figure}

In the final experiment the response of the irradiated and annealed sample was measured at the several fixed photon energies available at the beamline. Under the conditions of negligible re-scattering and self-absorption of luminescence, which are applicable to the thin diamond plate of the optical grade, the response as a function of the X-ray photon energy $E_X$ can be approximated with \cite{Rogalev02}:
\begin{equation}
R(E_X) \propto F_0 E_X (1 - \exp(-\mu(E_X)t)),
\label{eq:R}
\end{equation}
where $F_0$ is the incident photon flux, $\mu(E_X)$ is the X-ray attenuation coefficient of the material (for practical purposes mass-energy attenuation coefficient is often used), and $t$ is the thickness of the plate. Here it is assumed that the energy conversion efficiency and the quantum yield do not depend on the X-ray photon energy, which is a valid assumption for hard X-rays since the photon energies are substantially above the characteristic energies of any electronic transitions of the carbon atom. Figure~\ref{fig:r_en} shows the measured luminescence response normalized by $F_0$ and $E_X$ as a function of the photon energy. This ratio should be proportional to the absorptivity of the diamond plate according to Eq.~\ref{eq:R}. The absorptivity of the sample for a 100-$\mu$m-thick plate, scaled to the experimental data using the optimal proportionality coefficient is plotted with a solid line. The agreement between the experiment and Eq.~\ref{eq:R} is good. The observed minor discrepancies can be attributed to uncertainties in determination of the incident photon flux. 

\section{Summary}
In summary, we have demonstrated a more than one order of magnitude improvement in responsivity of X-ray excited luminescence in thin (nearly transparent for hard X-rays) diamond scintillator by introduction of NV defect states. The results of synchrotron X-ray measurements show that the new scintillator has a linear response to the incident hard X-ray flux density in the range from 7.6$\times 10^8$ to 1.26$\times 10^{12}$ photons/s/mm$^2$. This was demonstrated using a simple imaging scheme in the visible range with exposure times for individual frames from 0.01 to 5~s. Thus, an imaging device using the new scintillator can provide minimally invasive real-time transmission-mode imaging of photon flux density without disruption of X-ray instrument operation. 
Further improvement in responsivity can be achieved by exploring other highly luminescent defect centers, reducing influence of competing non-radiative processes (e.g., via control of crystal lattice quality) \cite{pat:Antipov19} as well as by using more advanced imaging detectors and image intensifiers. Future work could be extended to detailed studies of spatial resolution and time response of the new scintillator. Outcomes of such studies may lead to the next-generation technology for beam diagnostics at large scale X-ray user facilities.  
%
%\newpage
\acknowledgements

This work is based upon research conducted at the Cornell High Energy Synchrotron Source (CHESS) which is supported by the National Science Foundation under award DMR-1332208.
The work was supported by Department of Energy SBIR grant DE-SC0019628. 

%\bibliography{/home/oxygen/SSTOUPIN/Work2/Literature/bib/references}
%\bibliography{/nfs/chess/user/sas456/Work3/Literature/bib/references}

\end{document}